\documentclass{article}
\usepackage{graphicx}
\usepackage{epsfig}
\usepackage{amsfonts}
 \usepackage{amssymb}

\date{\today}

\newcommand{\insertplot}[5]{\begin{figure}
 \hfill\hbox to 0.05in{\vbox to #5in{\vfill
 \inputplot{#1}{#4}{#5}}\hfill}
 \hfill\vspace{-.1in}
 \caption{#2}\label{#3}
 \end{figure}}
 \newcommand{\inputplot}[3]{
 \special{ps: plotfile #1}
}
\newcounter{fig}

\newcommand{\ee}{\end{equation}}
\newcommand{\eea}{\end{eqnarray}}
\newcommand{\be}{\begin{equation}}
\newcommand{\bea}{\begin{eqnarray}}

\newcommand{\pont}{{\,^\ast\!}R\,R}

\begin{document}

 \title{ 
Remarks on the
Taub-NUT solution 
\\
in Chern-Simons modified gravity 
} 

\author{
{\large Yves Brihaye}$^{\dagger}$ 
{\large } 
and {\large Eugen Radu}$^{\ddagger}$  
\\ 
\\
$^{\dagger}${\small Physique-Math\'ematique, Universite de
Mons-Hainaut, Mons, Belgium}
\\
$^{\ddagger}${\small Departamento de F\'isica da Universidade de Aveiro and CIDMA,} 
\\
   {\small Campus de Santiago, 3810-183 Aveiro, Portugal}
}

\maketitle

\begin{abstract}  
We discuss the generalization of the NUT spacetime in General Relativity (GR)
within the framework of the (dynamical) Einstein--Chern-Simons (ECS) theory
with a massless scalar field. 
These configurations approach asymptotically the NUT spacetime and
 are characterized by the `electric'  and `magnetic' mass parameters and a scalar `charge'.
The solutions are found both
analytically and numerically. 
The analytical approach is perturbative around the Einstein gravity background.
Our results indicate that the ECS configurations
share all basic properties of the NUT spacetime in GR.
However, when considering  the  solutions inside the event horizon, 
we find that in contrast to the GR case,
the spacetime curvature grows (apparently) without bound. 

\end{abstract}

\section{Introduction}

The Einstein--Chern-Simons (ECS) theory \cite{Jackiw:2003pm}
is one of the most interesting generalizations of the
General Relativity (GR) \cite{Berti:2015itd}.
In its dynamical version, this model possesses a (real) scalar field $\phi$,
with an axionic-type coupling with the Pontryagin density \cite{Alexander:2009tp}.
As such, its  action contains extra-terms quadratic in the curvature
which can potentially lead to new effects in the strong-field regime.
Moreover, this model 
is motivated by string theory 
results
\cite{Campbell:1990fu}
 and  occurs also in the framework of
loop quantum gravity 
\cite{Taveras:2008yf},
\cite{Ashtekar:1988sw}.

In contrast to its Einstein--Gauss-Bonnet counterpart (in which case 
$\phi$ couples to the Gauss-Bonnet scalar),
it can be shown that any static spherically symmetric solution of GR is also a solution
of ECS gravity.
Therefore this model  is almost unique,
 as it leads to different results 
only in the presence of a parity-odd source such as rotation.
However, despite
the presence  in the literature  of some partial results
\cite{Cambiaso:2010un},
\cite{Cambiaso:2010cg},
\cite{Konno:2014qua}, 
the generalizations of the (astrophysically relevant) Kerr solution 
in ECS theory is still unknown,
presumably due to the complexity of the problem.
Therefore the  study of ECS
 generalizations of known GR rotating solutions
is a pertinent task which, ultimately,
could lead to some progress in the Kerr problem.

\medskip

One of the most intriguing solutions  of GR has been found in 1963
by  Newman, Tamburino and Unti (NUT)
\cite{NUT}.
This is a generalization of the 
Schwarzschild solution which 
solves the Einstein vacuum field equations,
possessing in addition to the mass parameter $M$
an extra-parameter--the NUT charge $n$.
In its  usual  interpretation, it describes a gravitational
dyon with both ordinary and magnetic mass.
The NUT charge $n$ plays a dual role to ordinary ADM mass $M$, 
in the same way that electric 
and magnetic charges are dual within Maxwell theory \cite{dam}.
This solution has a number of unusual
properties, becoming renowned for being 
{\it `a counter-example to almost anything'} 
\cite{misner-book}.
For example, 
the NUT spacetime is not asymptotically flat in the usual sense
although it does obey the required fall-off conditions,
and, 
moreover, contains closed timelike curves.
As such, it is cannot be taken as 
  a realistic model for a macroscopic object,
although its Euclideanized version
might play a role in the context of quantum gravity
\cite{Hawking:ig}.

\medskip

For the purposes of this work,
the NUT metric is interesting from another point of view:
its  line-element can be taken as Kerr-like,
 in the sense that it has a crossed 
 metric component $g_{\varphi t}$, see (\ref{metric}) bellow.
This term does not produce an ergoregion but it leads to
an effect similar to the dragging of inertial
frames \cite{Zimmerman:kv}.
Moreover,
one can say that a NUT spacetime consists of two counter-rotating regions,
with a vanishing total angular momentum
\cite{Manko:2005nm},
\cite{Kleihaus:2013yaa}.
Therefore,
 the study of its generalization in the framework
of ECS theory  is a legitimate task.

Also, one should mention that 
 the NUT solution has been generalized already in various models.
For example, nutty solutions with gauge fields have been
has been found in 
\cite{Brill},
\cite{Radu:2002hf},
\cite{Brihaye:2005ak}.
The low-energy  string theory
possess also nontrivial solutions
with NUT charge  (see $e.g.$ \cite{Johnson:1994ek}).

\medskip

The  paper is structured as follows: in the next Section we 
review the basic framework of the model which includes the metric and scalar field
Ansatz.
Some  properties of general nutty
solutions are also discussed there. 
In Section 3  
we present the results of a perturbative construction of solutions
as a power series in the CS coupling constant. 
The basic properties of the non-perturbative
configurations   are discussed in Section 4.
We conclude with Section 5 where the results are compiled.
There we present also our results for 
 the Taub region of the solutions
and give arguments that the solution is divergent there.

\section{The framework}

\subsection{The Chern-Simons modified gravity}

The action of the dynamical CS modified gravity is
provided by
\begin{eqnarray}
\label{CSaction}
I=\int d^4x  \sqrt{-g} 
\left(
 \kappa R+ \frac{\alpha}{4} \phi
{{\,^\ast\!}R\,R} 
 -\frac{1}{2} g^{ab} (\nabla_a \phi) (\nabla_b \phi) 
-V(\phi)
\right) ~,
\end{eqnarray}
where $g$ is the determinant of the metric $g_{\mu\nu}$, 
$R$ is the Ricci scalar 
and we note $\kappa^{-1} = 16 \pi G$.
 The quantity $\pont$ is the Pontryagin density, defined via
\begin{eqnarray}
\label{pontryagindef}
\pont=  
 {\,^\ast\!}R^a{}_b{}^{cd} R^b{}_{acd}~,~~{\rm with}
~~{^\ast}R^a{}_b{}^{cd}=\frac12 \epsilon^{cdef}R^a{}_{bef}~,
\end{eqnarray}
(where $\epsilon^{cdef}$ is the 4-dimensional Levi-Civita tensor). 
The gravity equations for this model read 
\begin{eqnarray}
\label{ECSeqs}
R_{ab}-\frac12 g_{ab}R = \frac{1}{2\kappa}T_{ab}^{(eff)},~~ {\rm with}~~ T_{ab}^{(eff)}=  T_{ab}^{(\phi)}-2 \alpha  C_{ab}  ,
\end{eqnarray}
where  
\begin{eqnarray}
\label{Ctensor}
C^{ab} = (\nabla_c \phi)
\epsilon^{cde(a}\nabla_eR^{b)}{}_d+
(\nabla_{c}\nabla_{d} \phi)
{\,^\ast\!}R^{d(ab)c}\,,   
\end{eqnarray}
and
$T_{ab}^{(\phi)} $ is the energy-momentum tensor of the scalar field, 
\be
\label{Tab-theta}
T_{ab}^{(\phi)}
=    \left(\nabla_{a} \phi \right) \left(\nabla_{b} \phi \right) 
    - \left [ \frac{1}{2}  g_{a b}\left(\nabla_{c} \phi \right) \left(\nabla^{c} \phi\right) + g_{ab}  V(\phi) \right]~.
\ee
The scalar field solves the Klein-Gordon equation in the presence of a source term given by the Pontryagin density,
\be 
\label{KGeq}
\nabla^2 \phi = \frac{dV}{d\phi} - \frac{\alpha}{4} \pont.
\ee

To simplify the picture, in this work we shall report results for a massless, non-selfinteracting scalar
only, $V(\phi)=0$.

\subsection{The Ansatz }
We consider a NUT-charged spacetime whose metric can be written locally in the form
\begin{eqnarray}
\label{metric}
ds^2=\frac{dr^2}{N(r)}+g(r)(d\theta^2+\sin^2 \theta d\varphi^2)-N(r)\sigma^2(r)(dt+4n\sin^2 \frac{\theta}{2} d\varphi)^2~,
\end{eqnarray} 
while the scalar field depends on the $r$-coordinate only, $\phi=\phi(r)$.
Here $\theta$ and $\varphi$ are the standard angles parametrizing 
an $S^2$ with the usual range.
As usual,
we define the NUT parameter\footnote{
One should remark that $n$ should be viewed as an input parameter of the model,
similar $e.g.$ to the cosmological constant in Einstein gravity.}
 $n$
(with $n\geq 0$, without any loss of generality),
in terms of the coefficient appearing in the differential
$dt+4n\sin^2 \frac{\theta}{2} d\varphi$.

 The form of $N(r),~\sigma(r)$ and $g(r)$ emerges as result of demanding 
the metric to be a solution of the ECS equations (\ref{ECSeqs})
(note the existence of a metric gauge freedom in (\ref{metric}), which is fixed later by convenience).
The equations satisfied by these functions (and the corresponding one
for $\phi(r))$ 
are rather complicated and we shall not not include them here.
However, we notice that they can also be derived from the effective action
\begin{eqnarray}
\label{Leff}
{\cal L}_{eff}={\cal L}_E+\kappa 
\left(
\frac{\alpha}{4}{\cal L}_{CS}+{\cal L}_\phi
\right)~,
\end{eqnarray} 
where
\begin{eqnarray}
\label{LE}
\nonumber
&&
 {\cal L}_E=
2\sigma
\left[
1+\left(\frac{N'}{2N}+\frac{g'}{4g}+\frac{\sigma'}{\sigma} \right)Ng'+\frac{\sigma^2 N}{g}n^2
\right],
~~~~
{\cal L}_{\phi}=-\frac{1}{2}N\sigma g \phi'^2~,
\\
\label{LCS}
&&
\nonumber 
{\cal L}_{CS}=
8n\frac{N\sigma^2}{g}
\left[
(\frac{N'}{N}-\frac{g'}{g}+\frac{2\sigma'}{\sigma})(1+4n^2\frac{N\sigma^2}{g} )\phi
+         \left(
\frac{N'^2}{4N^2}+\frac{g'^2}{4g^2}+\frac{\sigma'^2}{\sigma^2}
 -\frac{g'N'}{2gN}-\frac{g'\sigma'}{g\sigma}+\frac{N'\sigma'}{N\sigma}
         \right)Ng \phi'
				\right],
\end{eqnarray}
(where a prime denotes a derivative $w.r.t.$ the radial coordinate $r$).
Remarkably, one can see that,
due to the factorization of the angular dependence for the
 metric Ansatz (\ref{metric}),
 all functions solve
$second$ order equations of motion\footnote{Without this factorization, the metric functions
would
solve third order partial differential equations, 
this being $e.g.$ the case of the Kerr metric in ECS theory. }.

The reduced action (\ref{Leff}) makes transparent the scaling symmetries of the problem.
For example, to simplify the analysis, it is convenient to
work with conventions where $\kappa=1$
(this is obtained by rescaling the scalar field and the coupling constant $\alpha$).
Then the system  still has a residual scaling symmetry
\begin{eqnarray}
\label{scaling}
\alpha \to \alpha \lambda^2,~~
r\to \lambda r,~~
n\to \lambda n,~~
{\rm and}~~~
g\to \lambda^2 g,
\end{eqnarray}
which can be used to fix the value of $\alpha$ or $n$.

Finally, we note that
the NUT solution is found for $\alpha=0$, $\phi=0$,
being  usually written for a gauge choice with
\begin{eqnarray}
\label{TN}
\sigma(r)=1 ~~{\rm and}~~~N(r)=1-\frac{2(M r+n^2)}{r^2+n^2},~~~g(r)=r^2+n^2,
\end{eqnarray}
possessing a nonvanishing Pontryagin density
\begin{eqnarray}
\label{TN-RR}
 {{\,^\ast\!}R\,R}=\frac{96n^2}{(r^2+n^2)^6}\left(n^2(n^2-3r^2)+M r(3n^2-r^2) \right) \left(n^2(M-3r)+r^2(r-3M)\right),
\end{eqnarray}
(and thus it cannot be promoted to a solution of the ECS model).
This metric  has an (outer) horizon located at\footnote{Note that,
different from the case of a Schwarzschild black hole,
 a negative value  of the 'electric' mass $M$ is 
allowed for the NUT solution. 
Such configurations are found for $0<r_H<n$ and do not possess a Schwarzschild limit.}
\begin{eqnarray}
\label{rH}
 r_H=M+\sqrt{M^2+n^2}>0.
\end{eqnarray} 
Here, similar to the Schwarzschild limit, $N(r_H) = 0$ 
is only a coordinate singularity where all curvature invariants are finite. 
In fact, a nonsingular extension across this null surface can be found just 
as at the event horizon of a black hole.

\subsection{General properties}

Some basic properties of the line element (\ref{metric})
are generic, independent on the specific details of the considered gravity model.
As a result,  the general nutty configurations always share  the same troubles 
exhibited by the original NUT solution in GR.
For example,  
the Killing symmetries of (\ref{metric})
are time translation 
and $SO(3)$ rotations.
However, spherical symmetry in a conventional sense is lost,
 since the 
rotations act on the time coordinate as well. 
Moreover,  for $n\neq 0$,
the metric (\ref{metric}) has a singular symmetry axis.
However, following the discussion in \cite{misner-book} for the GR limit,
these singularities can be removed by appropriate identifications and changes in 
the topology of the spacetime manifold, which imply a periodic time coordinate.
Then such a configuration 
 $cannot$ be interpreted properly as black hole.
In fact, 
 the pathology of closed timelike curves  is not 
special to the NUT solution in GR 
but afflicts all solutions with a "dual" magnetic mass in general  \cite{Magnon}.
As discussed in \cite{Mueller:1986ij}, 
this condition emerges 
only from the asymptotic form of the fields.
Therefore, it is not sensitive to the precise details of the nature of the source, 
or the precise nature of the theory of gravity at short distances.

\medskip

In our approach we are interested in 
solutions whose far field asymptotics
are similar,  to leading order,
to those of the Einstein gravity solution (\ref{TN}),
with $N(r)\to 1$, $g(r)\to r^2$, $\sigma(r)\to 1$
and $\phi(r)\to 0$ as $r\to \infty$.
The solution will posses also an horizon at $r=r_H>0$,
where
$N(r_H)=0$, and $g(r)$, $\sigma(r)$ strictly positive.
   
In the absence of a global Cauchy surface,
 the  thermodynamical description of (Lorentzian signature) nutty solutions
is still poorly understood.
However, one can still define a temperature of solutions via the
surface gravity associated with the Killing vector $\partial/\partial t$,
\begin{eqnarray}
T_H=\frac{1}{4\pi}N'(r_H)\sigma(r_H),
\end{eqnarray}
and also an even horizon area \cite{Pradhan:2014tya}
\begin{eqnarray}
A_H=\int_0^\pi d\theta \int_0^{2\pi}d \varphi \sqrt{g_{\theta \theta} g_{\varphi \varphi}} \big |_{r=r_H}
=4\pi g(r_H).
\end{eqnarray}
The  mass of the 
solutions can be computed by employing 
the quasilocal formalism in conjuction with the boundary counterterm method
\cite{Astefanesei:2006zd}.
A direct computation shows that, similar to the Einstein gravity case,
the mass of the solutions
is identified with the constant $M$ in the far field expansion of
the metric function  $g_{tt}$,
\begin{eqnarray}
g_{tt}=-1+\frac{2M}{r}+\dots~.
\end{eqnarray} 

\section{A perturbative approach}
An exact solution of the equations 
(\ref{ECSeqs}), (\ref{KGeq})
 can be found in the limit of small $\alpha$, by treating
the ECS configurations as perturbations around the Einstein gravity background. 
Here we have found
convenient to work in a gauge with
\begin{eqnarray}
\label{pert1}
g(r)=r^2+n^2~.
\end{eqnarray} 
Then we consider a perturbative Ansatz with
\begin{eqnarray}
\label{pert11}
N(r)=N_0(r)(1+\alpha^2N_2(r)+\dots),~~
 \sigma(r)=1+\alpha^2\sigma_2(r)+\dots,~~
\phi(r)=\alpha \phi_1(r)+ \dots,~~
 \end{eqnarray} 
where $N_0=1-2(M_0 r+n^2)/{(r^2+n^2)}$ corresponds to the solution in Einstein gravity.  
 
To this order, one arrives at the following 
system of linear ordinary differential equations
\begin{eqnarray}
&&
\nonumber
\label{eqN2}
rN_2'+\frac{1}{N_0}N_2-\frac{6n^2}{g}\sigma_2
 =
\frac{2n}{g^2}
\bigg(r(r^2-3n^2)+M_0(n^2-3r^2)  \bigg)
\left(
\phi_1''-\frac{r(r^2-3n^2)+M_0(n^2-3r^2)  }{N_0g^2}\phi'
\right)
-\frac{1}{4}g\phi_1'^2, 
\\
\label{eqs2}
&&
r\sigma_2'+\frac{2n^2}{g}\sigma_2=\frac{1}{4}g\phi_1'^2
-\frac{n}{g^2}\bigg(r(r^2-3n^2)+M_0(n^2-3r^2)  \bigg)\phi_1'',
\\
\nonumber
\label{eqphi1}
&&
\phi_1''-\frac{2(M_0-r)}{N_0g}\phi'=\frac{24n}{N_0 g^6}
\bigg(M_0 r(r^2-3n^2)-n^2(n^2-3r^2) \bigg) \bigg(r(r^2-3n^2)+M_0(n^2-3r^2)  \bigg).
\end{eqnarray} 
When solving them, there are four integration constants. 
These constants are
chosen such that the corrected  NUT metric is still smooth at 
$r = r_H$ and approaches
a background with
$N(r)\to 1$ and $\sigma(r)\to 1$ asymptotically,
while $\phi(r)\to 0$.
Then, to lowest order, the solution has the generic structure
\begin{eqnarray}
\label{pertgen}
{\cal F}=P_0(r)+P_1(r) \arctan\left(\frac{n}{r} \right)+P_2(r) \log \left(\frac{(n^2+r^2)r_H^2}{(n^2+r r_H)^2} \right)~,
\end{eqnarray} 
with ${\cal F}=\{N_2,\sigma_2,\phi_1 \}$.
The functions $P_0$, $P_1$ and $P_2$
are  ratio of polynomials,
possessing a simple form for  $\phi_1$ only,  
with
\begin{eqnarray}
\label{pertgen1}
&&
 P_0=\frac{n^2}{(r^2+n^2)^3}
\left(
\frac{(n^2-r_H^2)}{n r_H}(n^2+\frac{(r^2-n^2)^2}{4n^2})+4rn
\right)
-\frac{r}{2n(r^2+n^2)},
\\
\nonumber
&&
P_1=\frac{1}{n^2},~~P_2=-\frac{r}{2n(r^2+n^2)}-\frac{n^2-r_H^2}{4 n r_H},
\end{eqnarray} 
the corresponding expressions for $N_2,\sigma_2$
being too complicated to display here.
To this order in perturbation theory,
one finds to following far field expression of the scalar field 
\begin{eqnarray} 
\phi_1(r)=\frac{q}{r}-\frac{n(n^2-r_H^2)}{4r_H^3}\frac{1}{r^2}+\dots,
~~{\rm with}~~~q=\frac{n}{2r_H^2}>0,
\end{eqnarray}
while the mass parameter has the following expression
\begin{eqnarray} 
\label{mass-cor}
M=M_0+\alpha^2 M_2,~{\rm with}~M_2=\frac{1}{64 n^5 r_H^5} 
\left( 
U_0(n,r_H)+U_1(n,r_H)\arctan(\frac{n}{r_H})+U_2(n,r_H)\log(\frac{ r_H^2}{ n^2+ r_H^2})
\right),{~~~}
\end{eqnarray}
 where $M_0=(r_H^2-n^2)/(2 r_H)$,
 and
\begin{eqnarray} 
&&
\nonumber
U_0=\frac{n}{210}
\left (
 429 n^6+2716 n^4 r_H^2-2555n^2r_H^4-3570r_H^6 
\right),
\\
&&
\nonumber
U_1=-r_H(n^2+r_H^2)(11n^4+5r^2r_H^2-22r_H^4),~~
U_2=\frac{1}{n}( r_H^4-n^4)( 5r_H^4-n^4)~.
\end{eqnarray}
The same type of  expression is found for the temperature, with
\begin{eqnarray} 
&&
\nonumber
T_H=\frac{1}{4\pi r_H}
\bigg[
1+\frac{\alpha^2}{6720n^2 r_H^4  (n^2+ r_H^2)^2 }
\bigg(
n^2
(429 n^8+5951 n^6r_H^2+343 n^4r_H^4-3115n^2r_H^6-1680 r_H^8)
\\ 
&&{~~~~~~~~~~~~~~~~}
-210(n^2-r_H^2) (n^2+ r_H^2)^3
\big(11nr_H \arctan(\frac{n}{r_H})-(n^2-3r_H^2)\log(\frac{ r_H^2}{ n^2+ r_H^2})
\big)
\bigg)
\bigg]~.
\end{eqnarray}
An inspection of the (\ref{mass-cor}) shows that $M_2$
is a strictly negative quantity. 
However,  
the CS correction to $T_H$
has no definite sign.
For a given $n$, it is negative for small $r_H$ and becomes strictly positive for large enough $r_H$
(in particular for $r_H>n$).

This approach can be extended to higher order in $\alpha$. 
Unfortunately,  
the resulting equations 
are too complicated for an analytical treatment.
Although they can be solved numerically,
we have preferred to consider instead  a fully nonperturbative approach.

\section{Numerical results}
The nonperturbative solutions are constructed by solving
numerically the ECS eqs. (\ref{ECSeqs}), (\ref{KGeq}),
as a boundary value problem.
In this approach, it is convenient to employ 
the same metric gauge
as in Einstein gravity,  and take
$\sigma(r)=1$.
Then we consider solutions in the domain 
$r_H\leq r<\infty$ (with $r_H>0$),
smoothly interpolating between the following boundary
values:
$N(r_H)=0$, $g(r_H)=g_0>0$, $\phi(r_H)=\phi_0$
and
$N=1$, $g=r^2$, $\phi=0$ as $r\to \infty$.
\begin{figure}[ht!]
\begin{center}
{\label{c1}\includegraphics[width=8cm]{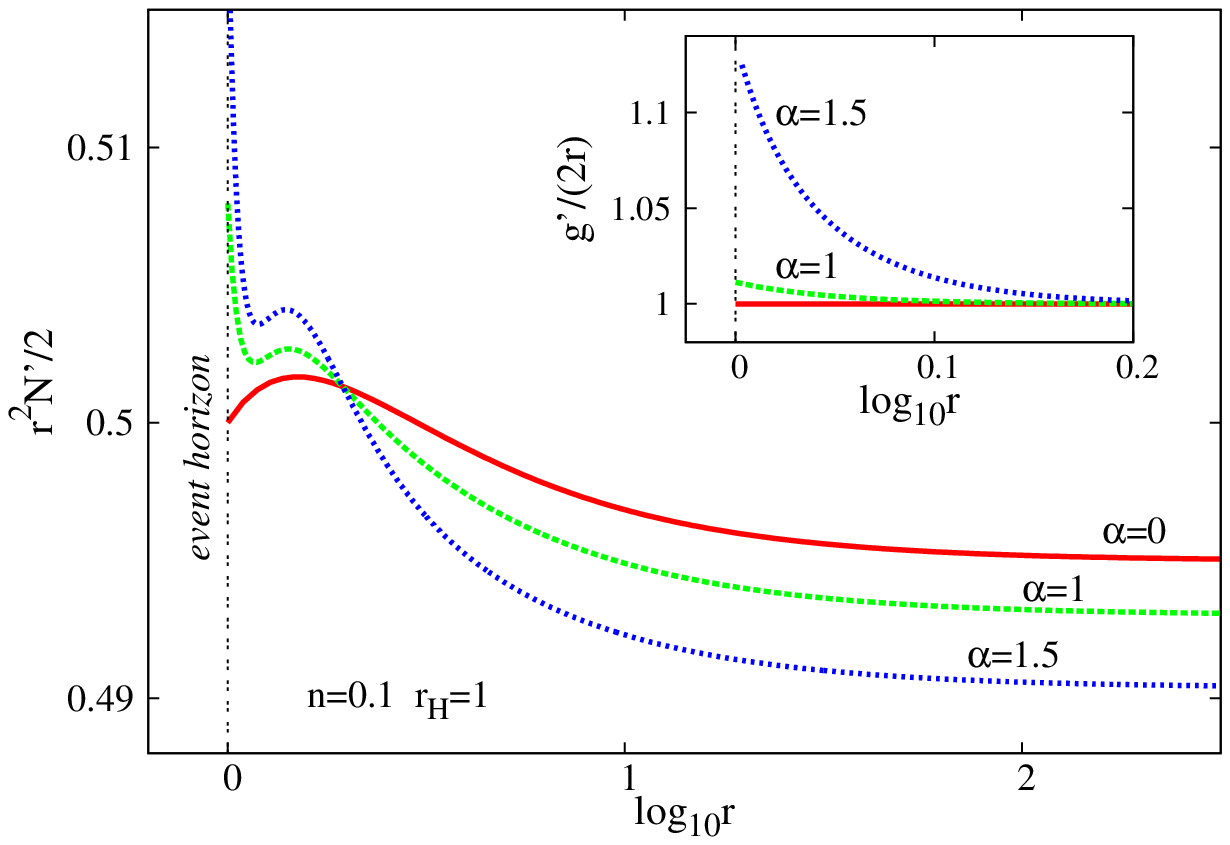}}
{\label{s0}\includegraphics[width=7.8cm]{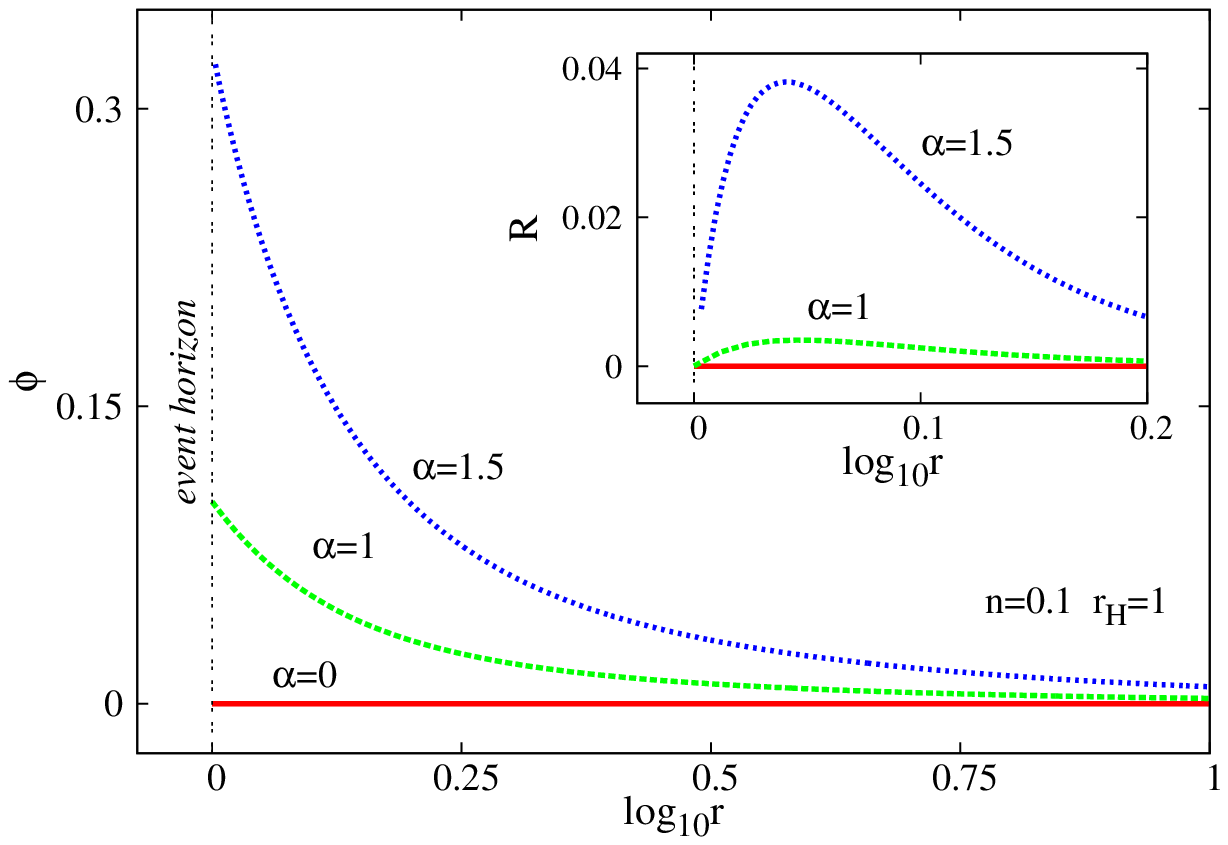}}
\caption{ {\it Left}: The profiles of $r^2 N'/2$ and $g'/(2r)$ are shown for several values of $\alpha$.
The solutions have $r_h=1$, $n=0.1$. 
{\it Right:}
The same for the scalar field $\phi$ and the Ricci scalar $R$.
\label{profile_Mass_0}
}
\end{center}
\end{figure} 
An approximate expression of the solutions compatible with these asymptotics
can easily be found. 
Its first terms as $r\to r_H$ are
\begin{eqnarray} 
&&
\label{hor}
N(r)= N_1(r-r_H)-\frac{1}{g_0}\frac{g_0^2+3N_1 n^2\alpha^2}{g_0^2-3N_1 n^2\alpha^2}(r-r_H)^2+\dots,
\\
\nonumber
&&
g(r)=g_0+\frac{1}{N_1} \frac{2g_0^2 }{g_0^2-3N_1 n^2\alpha^2}(r-r_H)+\dots,
~~
\phi(r)= \phi_0- \frac{6n \alpha }{g_0^2-3N_1 n^2\alpha^2}(r-r_H)+\dots,
\end{eqnarray}
$\{N_1,g_0,\phi_0 \}$ being three undetermined parameters,
while  
the leading order expansion in the far field is 
\begin{eqnarray} 
\nonumber
&&
N(r)=1-\frac{2M}{r}-\frac{2n^2}{r^2}+2M(n^2-\frac{q^2}{12})\frac{1}{r^3}+\dots,
\\
\label{inf}
&&
g(r)=r^2+(n^2-\frac{q^2}{4})-\frac{Mq^2}{3r}-\frac{q}{6}(3M^2q+n(nq-2\alpha))\frac{1}{r^2}+\dots,
\\
\nonumber
&&
\phi(r)=\frac{q}{r}+\frac{Mq}{r^2}+ (4M^2+n^2+\frac{q^2}{4})\frac{q}{3r^3}+\dots,
\end{eqnarray}
 containing
 the parameters
$M$ and $q$  fixed by numerics.
These constants are identified with the mass and the scalar `charge' of the solutions.
 
The ECS  equations have been solved by using a  solver 
which employs a Newton-Raphson method 
with an adaptive mesh selection procedure \cite{COLSYS},
the input parameters being $\{ r_H, n; \alpha \}$.
Starting with the GR solutions and slowly increasing $\alpha$, 
 we have found numerical evidence that the NUT metric possesses
non-perturbative generalizations in ECS theory.
For all considered solutions,
the metric functions $N(r),$ $g(r)$ are qualitatively very similar to their 
$\alpha=0$ counterparts,
while the scalar field  smoothly interpolate\footnote{Note that we could not find any indication for the
existence of excited solutions, the scalar field being always nodeless.} 
between the 
asymptotic expansions (\ref{hor}),  (\ref{inf}).
To reveal the effects of the CS term, 
we show in  Figure \ref{profile_Mass_0} (left) 
the function $r^2N'/2$ (whose asymptotic value corresponds to the mass $M$)
together with the function $g'/(2r)$ (whose values is one in GR).
The corresponding scalar field $\phi$ and the Ricci scalar $R$ 
are shown on the right hand panel of the figure. 
The solutions there have $r_H=1$, $n=0.1$ and several values of $\alpha$.

\begin{figure}[ht!]
\begin{center}
{\label{c2}\includegraphics[width=8cm]{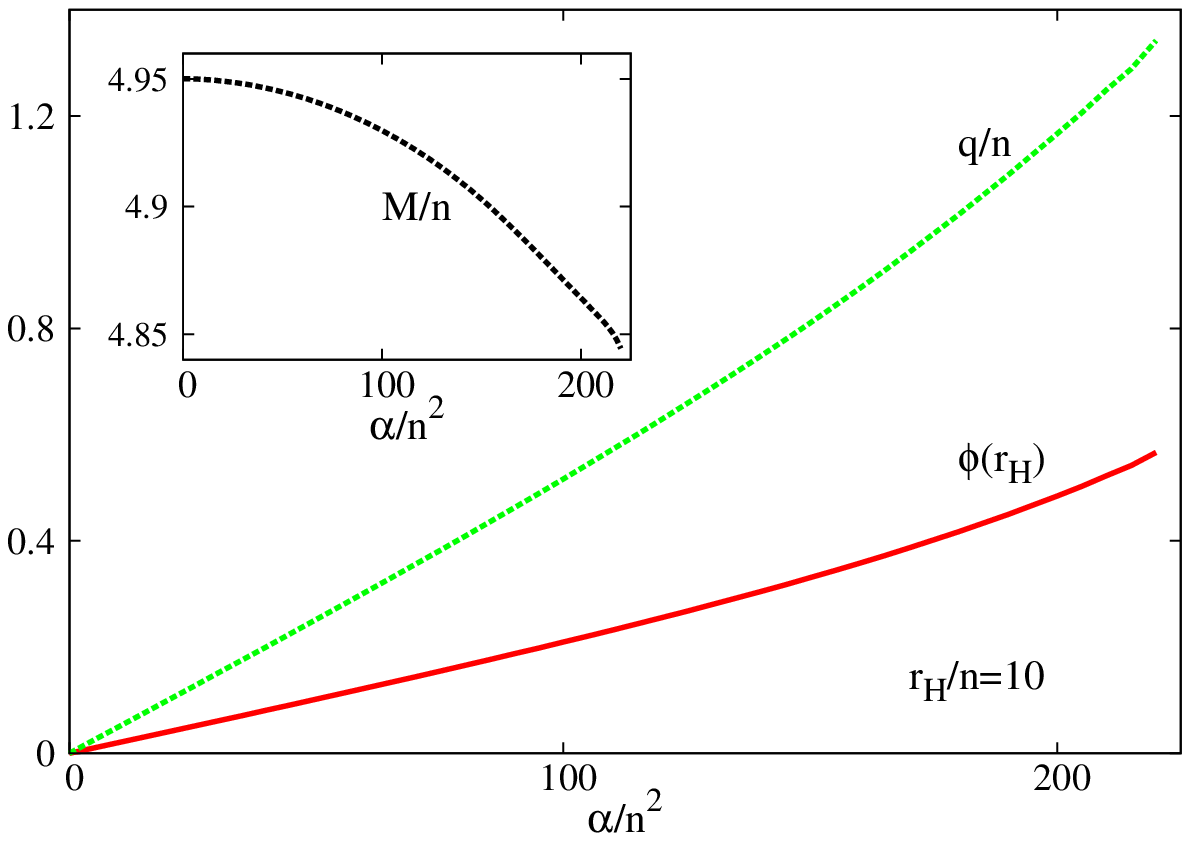}}
{\label{ss0}\includegraphics[width=7.8cm]{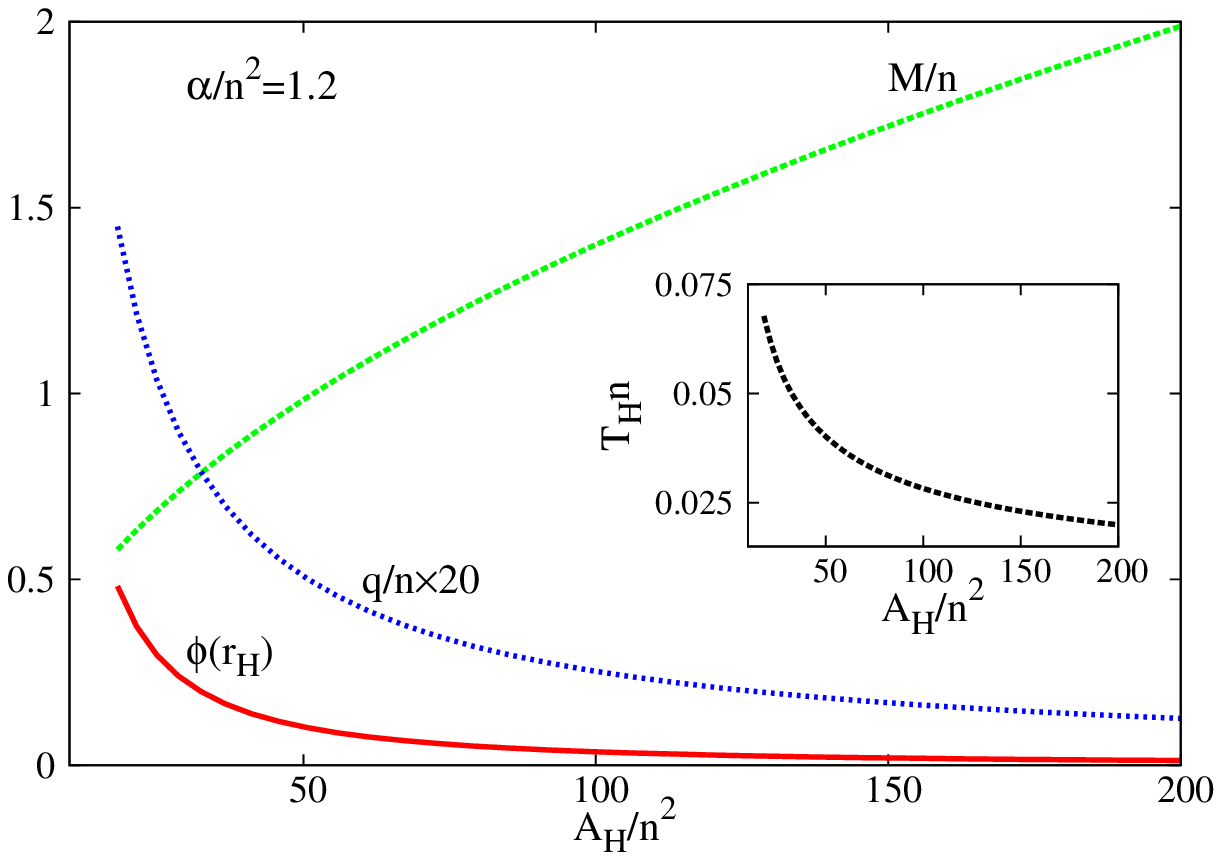}}
\caption{Left: Some parameters of the ECS solutions are shown as a function of  $\alpha$
(left) and of the horizon area (right).   
\label{data_Mass_0}
}
\end{center}
\end{figure}

The determination of the domain of existence of the solutions
would be a complicated task. 
In this work we will only 
report partial results in this direction, 
by analyzing the pattern of several classes of solutions  only.
Typical results of the numerical integration are shown\footnote{
The results in Figure  \ref{data_Mass_0} 
are likely to be generic, a (qualitatively) similar picture
being found for other values of the input parameters. }
 in Figure \ref{data_Mass_0}
as a function of $\alpha$ (left) and for a varying horizon size (right).
Note that all displayed quantities are expressed in units set by the NUT charge $n$,
being invariant under the transformation (\ref{scaling}).

As stated above, the ECS solutions smoothly emerge from the $\alpha=0$ GR ones. 
At the same time, the numerical results suggest that,
for given $(r_H,n)$,
 the value of the parameter $\alpha$
cannot be arbitrary large.
It turns out that, when the Chern-Simons
parameter becomes too large, the scalar field becomes very peaked at the horizon, with large values of
the Ricci scalar there, and
the overall numerical accuracy strongly decreases. 
Also, in agreement with the perturbation theory results, the mass $M$
decreases with $\alpha$, while the scalar `charge' $q$ is strictly positive,
increasing with $\alpha$.

When varying instead the horizon size for fixed $\{\alpha; n\}$  
(Figure \ref{data_Mass_0} (right)),
we notice the existence of a minimal value of  $A_H$, 
a feature shared with the GR solution.
For a given $n$,
this minimal value decreases as $\alpha$ increases.
Also, the scalar field vanishes gradually for large size of the horizon and becomes peaked at the horizon
as the minimal $A_H$
is approached.

\section{Further remarks. The issue of Taub solution}

The main purpose of this work was to investigate the basic properties of the Lorentzian NUT solution
in Einstein--Chern-Simons (ECS) theory, 
viewed as a 
toy model for a rotating configuration. 
Even if the primary interest is in the ECS generalization of the Kerr metric 
(which  would possess usual asymptotics and no causal pathologies), 
we hope
that, by widening the context to solutions with NUT charge, 
one may achieve a deeper appreciation of the model.

The problem has been approached from two different directions:
using an expansion in powers of $\alpha$
(the CS coupling constant) around  the GR solution,
 and solving the problem numerically.
As expected,  our results indicate that the basic properties (in particular 
the pathologies)
of the NUT solution persist for ECS configurations, 
without spectacular new features.
One interesting aspect which deserves further investigation
is the possible existence of a maximal value of $\alpha$, 
as suggested by the numerical results.

\begin{figure}[ht!]
\begin{center}
{\label{c3}\includegraphics[width=8cm]{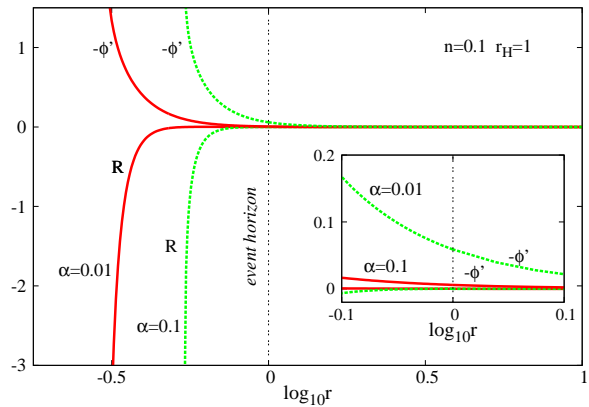}}  
\caption{The Ricci scalar $R$ and the derivative of the scalar field $\phi'$ are shown as a function
of $r$, inside and outside the horizon,
for two values of $\alpha$ and $r_H=1$, $n = 0.1$. 
\label{interior}
}
\end{center}
\end{figure} 

\medskip

This work can be continued in various directions.
For example,
once the geometry is known, one can study  
the effects of the CS term on the geodesic motion.
In the GR limit, $\alpha=0$, this problem has been extensively discussed in the literature,
 see $e.g.$
\cite{Zimmerman:kv},
\cite{LyndenBell:1996xj}-\cite{Jefremov:2016dpi}.
Restricting to null circular orbits, 
one can shown that,  for $\sigma(r)=1$,
 the radius $r=r_0>r_H$ of the photon sphere  
is a solution of the equation
\begin{eqnarray}
\label{eqn}
(N'g-Ng')|_{r=r_0}=0,
\end{eqnarray}
which in the GR case, reduces to
$r_0^3-3Mr_0^2-3n^2 r_0+M n^2=0$.
For $\alpha \neq 0$, the solution of (\ref{eqn})
is found numerically.
Our results indicate that  for a given $n$, the ratio 
$r_c/M$ increases with $\alpha$ (although 
for all solutions we have considered from this direction,
the differences $w.r.t.$ the GR case are at the level of a few percents).
It would be interesting to extend this study and to compute 
$e.g.$
the shadow  of the ECS
solutions.

\medskip

Returning to the GR solution
(\ref{TN}),
one remarks that 
the NUT metric is interesting  from yet 
another point of view.
By
continuing it through its  horizon at $r=r_H$
one arrives in the Taub universe,   which may be interpreted as a homogeneous, non-isotropic
cosmology with an  $S^3$ spatial topology.
(In fact, as discussed by Misner in \cite{Misner}, the NUT spacetime  
can be  joined analytically to the Taub spacetime
as a single Taub-NUT spacetime.)  
 Whereas the 
Schwarzschild solution has a curvature singularity at $r = 0$, 
this is not the case for $n\neq 0$ and  
the radius coordinate in Taub-NUT (TN) solution
may range on the whole real axis.

Since the regularity of the TN solution over the whole space-time is somehow exceptional,
it is natural
to address the question of the behaviour of the ECS solutions inside the horizon. 
Starting again with a perturbative approach, 
we remark that the solution derived in Section 3 holds also for $r<r_H$. 
Then   
one can show that 
the corrections $N_2(r)$ and $\sigma_2(r)$ to the TN
solution diverge\footnote{However, 
note that $\phi_1(r)$ remains finite at $r=0$.} 
as $1/r^2$ as $r\to 0$.
As expected, this divergence manifests itself also in 
the curvature invariants,
leading to a divergent character of the solutions,
at least to  lowest order in perturbation theory.

A similar conclusion is reached when considering a non-perturbative construction
of solutions inside the horizon.
This is a feasible problem, since we have obtained already the solutions at $r= r_H$.
This set is used as initial data to integrate inwards, on an interval
$[r_I , r_H]$, by decreasing progressively $r_I$.
The  results of the numerical (non-perturbative) integration can be summarized as follows. 
For all values of the parameters which we have considered, the integration inside can be performed only
for $r \in ]r_c, r_H]$ with $0 < r_c < r_H$.
The minimal value $r_c$ depends on the choice of the parameters
$\{r_H, n; \alpha\}$.
In particular, the Ricci scalar increases considerably in the limit $r\to r_c$, as shown
by Figs. \ref{interior} (note that a similar picture holds for the Kretschmann invariant $K$). 
These  results strongly suggest that all ECS solutions present an essential singularity at
$r=r_c$. 
Unfortunately,
we failed to find an analytical argument explaining this feature.
However, inspecting the different
functions entering in the equations, it turns out  that, for the chosen metric gauge,
$|\phi'(r)|$  strongly increases as $r\to r_c$ (see Fig. \ref{interior}). 
This induces strong variations of 
the functions $g',g''$ and likely leads to the divergence of $R$ and $K$.  
Finally, let us stress that -in agreement with the perturbative analysis-
the critical radius $r_c$ decreases towards zero when $\alpha$ decreases.
At the same time, its value increases with $\alpha$.
Moreover, the existing results suggest that 
this critical value reaches the horizon radius,
$r_c\to r_H$, as the maximal value of $\alpha$ (noticed in the 
previous Section) is approached,
which would imply a singular horizon in that limit.
However, a clarification of these aspects seems to require
another parametrization of the problem
and possibly
a different numerical approach.

\medskip

One should mention that we have also 
constructed ECS solutions with a massive scalar field,
$V(\phi)=\mu^2 \phi^2/2$.
However, all qualitative features of the massless solutions 
are recovered in that case.
In particular, the solution inside the horizon still appears to possess a singularity
for a critical value of $r$.

Finally, we remark that it would be interesting to find how
a (dynamical) CS term affects the properties of the
Euclideanized Taub-NUT solution.

\medskip
\medskip

{\bf Acknowledgement}
\\ 
 E. R. acknowledges funding from the FCT-IF programme.
This work was also partially supported 
by  the  H2020-MSCA-RISE-2015 Grant No.  StronGrHEP-690904, 
and by the CIDMA project UID/MAT/04106/2013.  
 


 \end{document}